\documentclass[twocolumn]{revtex4}
\usepackage{amssymb}
\usepackage{latexsym}
\usepackage{epsfig}
\usepackage{color}
\usepackage{float}
\usepackage{graphicx}
\usepackage{epstopdf}
\usepackage{color}

\begin{document}
\title{Non-minimal coupling inspires the Dirac cosmological model}
\author{H. Moradpour$^1$\footnote{hn.moradpour@maragheh.ac.ir}, H. Shabani$^2$\footnote{h.shabani@phys.usb.ac.ir}, A. H. Ziaie$^1$\footnote{ah.ziaie@maragheh.ac.ir}, U. K. Sharma$^3$\footnote{sharma.umesh@gla.ac.in}}
\address{$^1$ Research Institute for Astronomy and Astrophysics of Maragha
(RIAAM), University of Maragheh, P.O. Box 55136-553, Maragheh,
Iran\\
$^2$ Physics Department, Faculty of Sciences, University of Sistan and Baluchestan, Zahedan, Iran\\
$^3$ Department of Mathematics, Institute of Applied Sciences and
Humanities, GLA University,
    Mathura-281406, Uttar Pradesh, India}
\date{\today}
\begin{abstract}
In the framework of the generalized Rastall theory
(GRT), we study the ability of a non-minimal coupling between
geometry and matter fields in order to provide a setting which
allows for a variable $G$ during the cosmic evolution. In this
regard, the compatibility of this theory with Dirac hypothesis on
the variations of $G$ is investigated, and additionally, the
possibility of obtaining the current accelerated universe is also
addressed. In summary, our study indicates that, in GRT, having in
hand the $G$ profile, one may find the corresponding non-minimal
coupling between the energy source and geometry and vise versa, in
a compatible way with the current accelerated universe.
\end{abstract}

%\pacs{04.50.Kd; 95.36.+x; 98.80.-k; 98.80.Jk}
% \keywords{Cosmology; $f(R,T)$ Gravity; Dark Energy; Dynamical Systems Approach; Modified Theories of Gravity.}
\maketitle
\section{Introduction}\label{Intr}
The idea that $G$ (the Newtonian gravitational coupling)
has probably experienced diverse values during the cosmic
evolution has many motivations. It began with Dirac's proposal~\cite{dir1,dir2,dir3} which states that, the ubiquitousness of certain large dimensionless numbers (LDN's), arising in combinations of physical constants and cosmological quantities~\cite{WZE} was not a coincidence but an outcome of an underlying relationship between them~\cite{BTB}. In his proposal, Dirac pointed out that the electrical force between proton and electron within a hydrogen atom i.e., $F_e=e^2/4\pi\epsilon_0r^2$ is a large number being 40 orders of magnitude greater than their gravitational force $F_G=Gm_pm_e/r^2$, i.e.,
\begin{eqnarray}
	{\rm LN}_1=\frac{F_e}{F_G}=\frac{e^2}{4\pi\epsilon_0Gm_pm_e}\approx10^{40},
\end{eqnarray}
where $m_e,e,m_p,\epsilon_0$ and $G$ are the mass and charge of electron, the proton mass, the vacuum permittivity and gravitational constant, respectively. On the other side, the ratio of the age of the universe and the time for light to traverse an electron is also nearly of the same size, i.e.,
\begin{eqnarray}
{\rm LN}_2=\frac{t}{e^2/4\pi\epsilon_0m_ec^3}\approx10^{40}.
\end{eqnarray}
Dirac then suggested that the above two quantities are equal. As a result of such a relationship, some of the fundamental constants cannot remain constant for ever since ${\rm LN}_2$ varies with the age of the universe. According to Dirac's hypothesis, atomic parameters cannot change with time and thus $G$ should change inversely with time, i.e., $G\propto t^{-1}$~\cite{CHK}, see also~\cite{DIRACREV} for recent reviews. Since the advent of this idea, it has led to interesting implications within theoretical physics, and has attracted a great deal of attention during the past decades~\cite{ras2,vin,sab,bap,bee,wu,deg,bar1,bar2,
bar3,mans,gaz,clif,bro,sol,uza1,uza2,smo,fri,les}. {\bf Moreover, it} has
even interesting power to justify baryogenesis~\cite{les}, the
current and primary accelerated universes~\cite{ell} and can
support the de Sitter spacetime~\cite{uza1,uza2}.
\par
In Newtonian gravity one is allowed to write an explicit time variation of $G$ without the need of satisfying any further constraint. However, the situation is different in GR as there are further constraints to be satisfied. Consider the Einstein field equation $G^{\mu}_{\,\nu}=8\pi GT^{\mu}_{\,\,\nu}$ with the assumption of $G=G(t)$ and $c\equiv1$. If one takes the covariant divergence of this equation the left hand side vanishes as a result of Bianchi identity. Then, if the ordinary energy-momentum conservation law (OCL) is assumed to hold, i.e., $T^{\mu}_{\,\,\nu;\mu}=0$, one finds that $G$ must be a constant with respect to spacetime coordinates, i.e., $\partial G/\partial x^\mu=0$ always. In this respect, GR does not allow for any variation in the gravitational coupling $G$ owing to the fact that the Einstein tensor is divergence free and the divergence of energy-momentum tensor is also zero. Hence, in the light of Dirac's proposal, some modifications of GR field equation are essential. This is because, if we simply let $G$ to be a variable then the OCL is violated~\cite{CanutoAdams}. In this respect, investigating the effects of a varying $G$ can be performed only through modified field equations along with modified conservation laws. From these arguments, one may intuitively imagine that a varying $G$ could contribute as a new degree of freedom within the OCL. As in GR, $G$ denotes mutual relation between geometry and matter fields, hence, variations of $G$ together with the
violation of OCL may be considered as a signal for the idea that another relation between geometry and matter fields may exist that connects their changes to each other. However, there are modifications of GR with a varying $G$ that respect the OCL such as Brans-Dicke theory, in which, the dynamical scalar field $\phi$ can be considered as the origin of gravitational coupling and thus it varies as $G\propto\frac{1}{\phi}$~\cite{11,12,13,14,bar2}.
\par
%In fact, accepting this idea and bearing the Bianchi identity in mind,
%one can easily deduce that, even in general relativity (GR),
%ordinary energy-momentum conservation law (OCL) is not satisfied
%whenever $G$ is not constant, a signal to this idea that the
%violation of OCL and the evolution of $G$ are not necessarily
%separate phenomena, and may have common origin and
%aspects. 

OCL, as one of the cornerstones of GR~\cite{pois}, is not
respected in all modified gravity theories, for example, it is
broken in the non-minimal curvature matter coupling
theories~\cite{od1,all,koi,bert,hark,car,boh,ras1,mor1,mora}.
Rastall gravity is a pioneering theory in this area~\cite{ras1} in
accordance with various observations~\cite{li,raw1,raw2,raw3,maj,arb,rah1,rah2,rah3,mor2,man,ortiz2020,shabooni2020}
and its corresponding cosmology avoids the age and entropy
problems arisen in the framework of the standard
cosmology~\cite{fab}. In fact, this theory can even provide a
better platform for describing the matter dominated era compared
to the Einstein theory~\cite{raw2}. A generalized form of
this theory allows us to relate the current and primary
accelerated universe to the ability of the spacetime to couple
with the energy-momentum sources, filling the background, and in
fact, introduces this coupling as a candidate for dark energy and
inflaton field~\cite{mor1}.

In addition to inflationary models powered by employing
varying $G$ theories \cite{ell}, there are also other models to
describe inflation without considering an
inflaton field~\cite{jos,mor1,wat,gam}. In Ref.~\cite{mor1}, it has been
shown that while the existence of an interaction between the
geometry and matter fields may model the primary and
current inflationary eras, it does not necessarily lead
to the break-down of OCL. In fact, if geometry has the ability of
being non-minimally coupled with the matter fields, then
this ability may support the primary inflationary era and
the current accelerated phase~\cite{mor1}. To obtain these
results, authors focus on the special case of $T^{\mu \nu}_{\ \
;\mu}=0$, and find out the form of non-minimal coupling
in each cosmic era.

The study of various non-minimal couplings can at least make us
familiar with their consequences and properties which may finally
lead to a better understanding of spacetime that helps us
provide better predictions about its behavior and nature. In GRT,
cosmological scenarios~\cite{jos,mor1,das,lin} imply the power of
non-minimal coupling in $i$) providing both singular and
non-singular universes, $ii$) describing the particle production
process, $iii$) avoiding the coincidence problem, and $iv$)
playing the role of dark energy (unlike the original
Rastall theory \cite{mor1,batis}). In this regard,
thermodynamically it has also been shown that the confusion in
defining energy and some of its outcomes which may lead to the OCL
generalization (or equivalently, the breakdown of OCL) could make
the cosmos dark~\cite{mor3,mor4}.

Since in Rastall gravity, the gravitational coupling is a
constant, but differs from those of GR and Newtonian gravity
(NG)~\cite{mor5,ras1}, Rastall theory (and indeed a mutual
non-minimal coupling between the geometry and matter fields in the
Rastall way) cannot provide a theoretical basis for the probable
variations of $G$ during the cosmic evolution. These
points will be reopened in more details in the next section.

Motivated by the above arguments, it is reasonable to $i$)
examine the ability of non-minimal coupling between geometry and
matter fields in producing a non-constant $G$, and also $ii$)
study the results of a non-constant $G$ in the framework of
Rastall theory. The latter is tried to be answered by some authors
in Ref.~\cite{ref}, by combining Rastall and Brans-Dicke theories
with each other. In the present study, the changes in $G$ is not
originated by the Rastall theory meaning that the first part is
still unsolved and debateable. We therefore focus on GRT to
describe the compatibility of a non-minimal coupling with Dirac's idea on evolution of $G$. Indeed, we are eager
to show that, at least phenomenologically, a non-minimal coupling
may itself change $G$ and play the role of dark energy.

The present work is then arranged as follows. In Sects.~\ref{sec2}
and ~\ref{sec3}, a brief review on the Rastall theory and its
generalization~\cite{mor1} has been provided, and some of their
predictions about the variations of $G$ are addressed.
Sect.~\ref{sec4} includes our survey on the possibility of
explaining a well-known Dirac cosmological model, previously
introduced by other authors, within the framework of GRT. To show
the ability of non-minimal coupling in satisfying Dirac hypothesis
and describing the cosmic evolution, simultaneously, a new model
is also introduced in Sect.~\ref{sec5}. Sect.~\ref{sec6} is
devoted to concluding remarks. Here, we use $c=\hbar=1$ units.
%%%%%%%%%%%%%%%%%%%%%%%%%%%%%%%%%%%%%%%%%%
\section{Rastall theory and a model for varying $G$}\label{sec2}
Originally, P. Rastall argued that the OCL may not be valid in a curved spacetime
leading to~\cite{ras1}

\begin{eqnarray}\label{r0}
T^{\mu \nu}_{\ \ ;\mu}\neq0,
\end{eqnarray}

\noindent in the non-flat spacetimes. From the mathematical point
of view, $T^{\mu \nu}_{\ \ ;\mu}$ is a ranked one tensor field written
as $T^{\mu \nu}_{\ \ ;\mu}=Q^{ ,\nu}$ where $Q$ is an unknown
scalar function found out from other parts of physics, mathematics
and observations~\cite{ras1}. Since $Q$ is a scalar and Rastall
hypothesis admits the violation of OCL in a curved spacetime
(where Ricci scalar is not always zero), therefore Ricci scalar, R, can be considered as a suitable suggestion for $Q$, and thus~\cite{ras1}

\begin{eqnarray}\label{r1}
T^{\mu \nu}_{\ \ ;\mu}=\lambda^{\prime} R^{;\nu},
\end{eqnarray}

\noindent where $\lambda^{\prime}$ is called the Rastall constant
parameter. Using the Bianchi identity, it is easy to get

\begin{eqnarray}\label{r2}
G_{\mu \nu}+\kappa^{\prime}\lambda^{\prime} g_{\mu
\nu}R=\kappa^{\prime} T_{\mu \nu},
\end{eqnarray}

\noindent which $\kappa^{\prime}$ is a constant
\cite{ras1} called the Rastall gravitational coupling constant.
Applying the Newtonian limit on this result, we obtain~\cite{ras1}

\begin{eqnarray}\label{k1}
\frac{\kappa^{\prime}}{4\kappa^{\prime}\lambda^{\prime}-1}\left(3\kappa^{\prime}\lambda^{\prime}-\frac{1}{2}\right)=\kappa_G,
\end{eqnarray}

\noindent where $\kappa_G\equiv4\pi G$. Hence, since
$\kappa^{\prime}$ and $\lambda^{\prime}$ are constants, $G$ should
also be a constant as well (the current value of $G$, namely
$G_0$, is proper option leading to
$\kappa_G\equiv\kappa_{G_0}=4\pi G_0$). {We therefore conclude that, since the left hand side of (\ref{k1}) is a constant then} a mutual non-minimal interaction between the geometry and matter fields {within the framework of original version of} Rastall
{gravity} does not support the varying $G$ theories.
{Eq.~(\ref{k1}) also reveals that the Rastall gravitational
coupling constant ($\kappa^{\prime}$) differs from that of GR
($2\kappa_G=8\pi G$) and only if $\lambda^\prime=0$ then they will
be equal.}

It is also useful to note that one may use Eq.~(\ref{r2}) in order
to introduce the generalized energy-momentum tensor
$\Theta_{\mu\nu}=T_{\mu\nu}-(\kappa^{\prime}\lambda^{\prime})/(4\kappa^{\prime}\lambda^{\prime}-1)Tg_{\mu\nu}$
which finally leads to the GR counterpart form of the Rastall
field equations, given as $G_{\mu
\nu}=\kappa^{\prime}\Theta_{\mu\nu}$. In this manner, although the
obtained field equations are similar to those of GR, their
solutions for $T_{\mu\nu}$ differ in general from those of
GR~\cite{mor4,dar}, a result confirmed by various observational
data, {see e.g.,}~\cite{li,mor2,dar} and references therein).
\par
One can also generalize the Rastall theory  by considering $\lambda^\prime\rightarrow\lambda$, where $\lambda$ is a varying parameter. Therefore Eq.~(\ref{r1}) is extended as follows~\cite{mor1}

\begin{eqnarray}\label{gr0}
T^{\mu\nu}_{\ \ \ ;\mu}=\left(\lambda R\right)^{;\nu},
\end{eqnarray}

\noindent which finally leads to

\begin{eqnarray}\label{gr1}
G_{\mu\nu}+\kappa\lambda g_{\mu\nu}R=\kappa T_{\mu \nu},
\end{eqnarray}

\noindent where $\kappa$ is again a constant {but $\lambda$ can change over time}. Using the trace of
Eq.~(\ref{gr1}), one can also rewrite this equation as

\begin{equation}\label{gr2}
G_{\mu\nu}+\tau Tg_{\mu\nu}=\kappa T_{\mu\nu},
\end{equation}

\noindent in which

\begin{equation}\label{gr3}
\tau=\frac{\kappa^2\lambda}{4\kappa\lambda-1}.
\end{equation}

\noindent Now, since $\kappa$ is constant, the covariant
derivative of Eq.~(\ref{gr2}) leads to

\begin{equation}\label{gr4}
\tau^{,\nu}T+\tau T^{,\nu}=\kappa T^{\nu\,\,\,;\mu}_{\,\,\,\mu},
\end{equation}

\noindent meaning that even if OCL is respected and until
$\tau\neq constant$ (or equally, $\lambda\neq constant$), the
non-minimal coupling affects the evolution of the energy-momentum
source and vice versa~\cite{mor1}. Therefore, unlike the Rastall
theory, OCL can be met in this framework even in the presence of
non-minimal coupling. In this regard, it is shown that, in the
framework of Eq.~(\ref{gr1}), even if OCL is met, the accelerated
universe can be explained under the shadow of $\lambda$ {without resorting to} a dark energy source~\cite{mor1}.

Now, considering the Newtonian limit (ignoring the pressure of
$T_{\mu\nu}$ and utilizing relation $R_{00}=\nabla^2 \phi$, in
which $\phi$ denotes the Newtonian potential~\cite{mor6}), one can
easily find

\begin{eqnarray}\label{k2}
\frac{\kappa}{4\kappa\lambda-1}\left(3\kappa\lambda-\frac{1}{2}\right)=\kappa_G.
\end{eqnarray}

\noindent Due to the similarity of Eqs.~(\ref{gr1})
and~(\ref{r2}), one could expect that the Newtonian limit of field
equations~(\ref{gr1}) is obtainable by replacing $\kappa^{\prime}$
and $\lambda^{\prime}$ with $\kappa$ and $\lambda$, respectively,
in Eq.~(\ref{k1}). Eq.~(\ref{k2}) also indicates that $G$ (or
equally $\kappa_G$) does not necessarily remain constant in this
theory. Therefore, this generalization of Rastall theory provides
a basis for theories including a varying
$G$~\cite{dir1,dir2,vin,sab,bap,bee,wu,mans,gaz,bar1,bar2,bar3,clif,uza1,uza2,smo,fri,les}.
In fact, this equation tells that a non-minimal coupling between
the geometry and matter fields can make $G$ variable~\cite{mot}
meaning that such coupling can be considered as a theoretical
basis for varying $G$ theories.
%%%%%%%%%%%%%%%%%%%%%%%%%%%%%%%%%%%%%%%%%%%%%%%%%%%
\section{Newtonian limit, a model for running $G$, and the value of $\kappa$}\label{sec3}
Now, using Eq.~(\ref{k2}), and following Ref.~\cite{mor1}, in
which $\kappa\lambda\equiv\beta=[4+\theta(1+z)^3]^{-1}$, where
$\theta$ is an unknown constant and $z$ denotes the redshift, one
can obtain

\begin{eqnarray}\label{v1}
\kappa_G=\frac{\kappa}{2}\left[1-\frac{2}{\theta(1+z)^3}\right],
\end{eqnarray}

\noindent finally leading to

\begin{eqnarray}\label{v11}
\kappa_G=\frac{\kappa}{2}\left[1-\frac{2}{\theta}\right]\equiv\kappa_{G_0},
\end{eqnarray}

\noindent and

\begin{eqnarray}\label{v12}
\kappa_G=\frac{\kappa}{2},
\end{eqnarray}

\noindent for $z\rightarrow0$ and $z\rightarrow\infty$,
respectively. Based on Ref.~\cite{mor1}, whenever $0<\theta\leq1/2$ (leading to $\beta>0$), the current accelerated universe is
explainable in the presence of OCL, and without considering a dark
energy-like source. Moreover, expression
$\beta=[4+\theta(1+z)^3]^{-1}$ is {present} in both of the matter
dominated era (MDE) and the current accelerated
universe~\cite{mor1}. Hence, Eq.~(\ref{v12}) can be considered as
the value of $G$ at the beginning of MDE whereas the value of
$\kappa$ is obtainable by using Eq.~(\ref{v11})

\begin{eqnarray}\label{v13}
\kappa=\frac{8\pi G_0}{1-\frac{2}{\theta}},
\end{eqnarray}

\noindent combined with Eq.~(\ref{v12}) to see that
{$\kappa$, and thus} $\kappa_G$ are negative at the
beginning of MDE. Therefore, in the model proposed in
Ref.~\cite{mor1} which still respects OCL in the framework
of~(\ref{gr1}), $G$ is not always positive during the cosmic
evolution. Negative values of $\kappa$ provide a setting for
baryonic matters to support traversable wormholes in the Rastall
framework ~\cite{mor5}. {Moreover, in the framework of GRT, it has been shown that negative values of $\kappa$ could have their own effects on matter perturbations and formation of structures in large scale universe~\cite{AHH2020}. In this regard, overdense and underdense regions in the universe could form periodically so that both large scale structures and voids could form as the universe evolves from MDE to present time. Also, emergence of structures in a class of alternative theories of gravity has been reported in~\cite{Lohiya1996}, where the authors considered a non-minimally coupled scalar field in addition to an induced negative gravitational constant and studied structure formation with repulsive gravitation on the large scale. In the framework of general scalar tensor-theories, a cosmological mechanism has been proposed in which it is possible for $G$ to change sign from a positive branch (attracting) to a negative branch (repulsive gravity) and vice versa~\cite{Nunez2019}. {It is also worth mentioning that} negative values of $G$ have previously been reported in some other approaches studying the variations of $G$~\cite{bar2,uza1,uza2}.}

\par {Beside the effects of repulsive gravity (represented by a universal negative coupling) on the evolution of perturbations and formation of structures, the study of possible consequences of $\kappa<0$ on the stability of the model is of particular importance. In this regard, from the viewpoint of perturbative analysis, the existence of a repulsive gravity phase in the evolution of the universe could lead to growing models with respect to scalar perturbations producing then, large inhomogeneities. Hence a repulsive phase may destroy homogeneity and in this sense it may be unstable~\cite{Batista2001}.  In~\cite{Star1981}, it has been discussed that a transition from positive gravitational coupling $G$ to negative one results in an instability, in such a way that, small deviations from isotropy and homogeneity within the gravitational field will grow unboundedly, leading to a true cosmological singularity at the boundary between gravity and anti gravity. Also, investigating classical stability of the model through dynamical system approach is of long-standing interest and significance. Work along this line has been carried out for a class of GRT models~\cite{Lin2020}, where the authors have shown that the eventual fate of the universe ends in late time attractors which are classically stable. However, investigating these issues for the present model needs a deeper analysis with more scrutiny and future studies will be reported elsewhere.} {Finally, we} note that, since $\dot{G}$ does not decrease with time for $0<\theta\leq1/2$ ($\dot{G}>0$ in this manner), this model does not respect the Dirac's hypothesis claiming that $G$ should decrease as a function of time~\cite{dir1,dir2,vin,bap}. Hence, more comprehensive non-minimal couplings are needed to provide settings for Dirac hypothesis and also to model the cosmic evolution without considering a mysterious fluid (dark energy), simultaneously.
\subsection{Another possibility}

In Ref.~\cite{das}, {choosing
$\lambda=(1+d_0H)/[3\kappa(w+1)]$}, in which $w\equiv p/\rho$
(where $p$ and $\rho$ denote the pressure and energy density of
the cosmic fluid, respectively), it has been shown that
non-singular cosmic evolution is obtainable in GRT. In this case
$d_0$ is a free parameter, and some outcomes of this proposal in
various cosmic eras have also been studied in Ref.~\cite{das}.
Accepting this proposal along with considering {the} unit $\kappa=8\pi
G_0$ and also {assuming} $G(H_0)=G_0$ (which helps us in finding
$d_0$), one easily reaches

\begin{eqnarray}\label{w1}
G(H)=G_0\frac{3(1-w)H_0-6H}{(1-3w)H_0-4H},
\end{eqnarray}

\noindent where $H_0$ is the current value of $H$ {and use has been made of Eq.~(\ref{k2}).}
%%%%%%%%%%%%%%%%%%%%%%%%%%%%%%%%%%%%%%%%%%%%%%%%%%%
\section{Dirac cosmological model}\label{sec4}
As in the present model there is no evolution equation for the variation of $G$ which is promoted as a dynamical field, one then has to impose a suitable ansatz on the behavior of this parameter. Based on Dirac hypothesis, $G$ should decrease with time{\bf, i.e, $G\propto t^{-1}$~\cite{CHK}}. In general, one may consider $G=G_0f$, in which $f$ is a decreasing
function of time~\cite{dir1,dir2,vin,bap,clif}), in order to
preserve Dirac hypothesis. Now, combining Eq.~(\ref{k2}) with
$\kappa=8\pi G_0\alpha$~\cite{raw2}, along with Eqs.~(\ref{gr0}) and (\ref{gr1}) for a flat FLRW universe, one finds

\begin{eqnarray}\label{ras1}
&&\gamma\equiv\lambda\kappa=\frac{f-\alpha}{4f-6\alpha},\\
&&3\int(\rho+p)\frac{da}{a}=\frac{1}{2\alpha}\Big[(f-3\alpha)\rho-3(f-\alpha) p\Big],\nonumber\\
&&H^2=\frac{1}{6}\Big[(3\alpha-f)\rho+3(f-\alpha)p\Big],\nonumber\\
&&q=-1-\frac{\dot{H}}{H^2}=-1+\frac{3\alpha(\rho+p)}{\rho(3\alpha-f)+3(f-\alpha)p},\nonumber
\end{eqnarray}

\noindent whenever a fluid with energy density $\rho$ and pressure
$p$ fills the background. {We note that $\gamma$ is a varying parameter and}, $q$ and $a$ denote
deceleration parameter and scale factor, respectively, and we also have assumed $8\pi G_0=1$.

\begin{figure}
\begin{center}
\includegraphics[width=8cm]{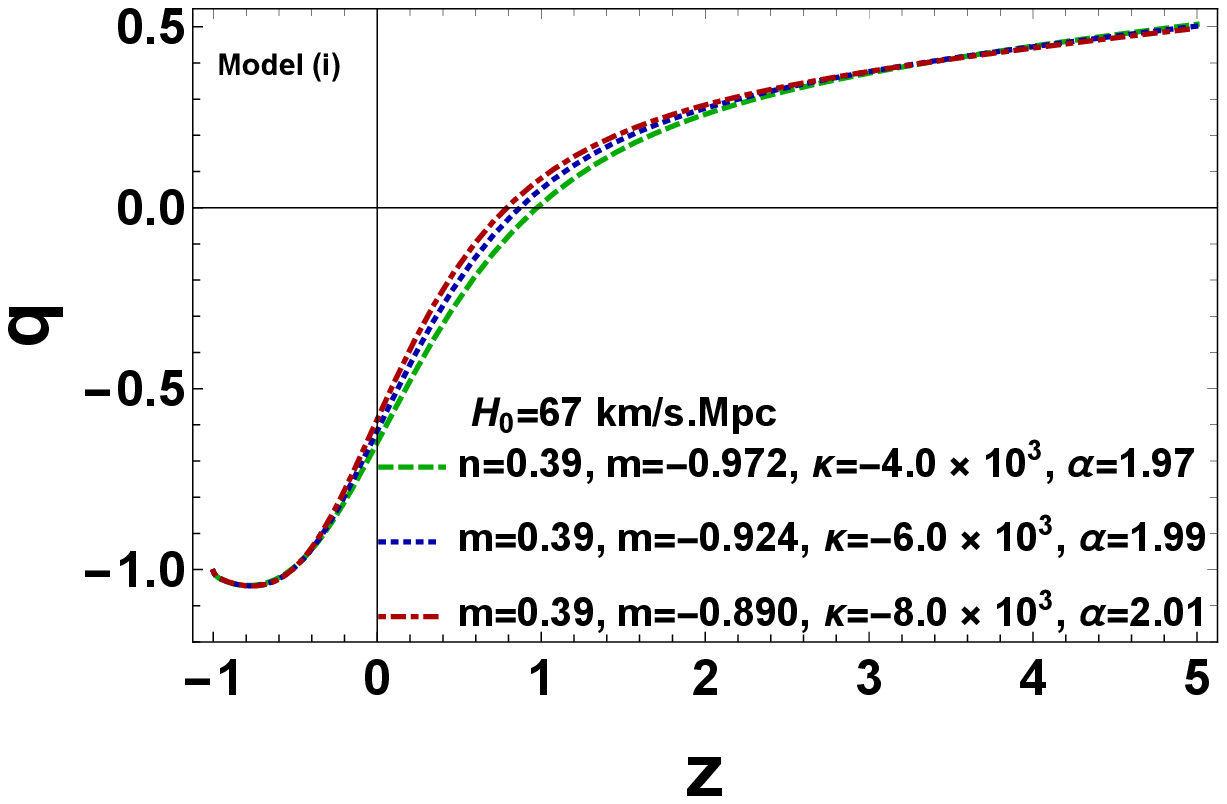}
\includegraphics[width=8cm]{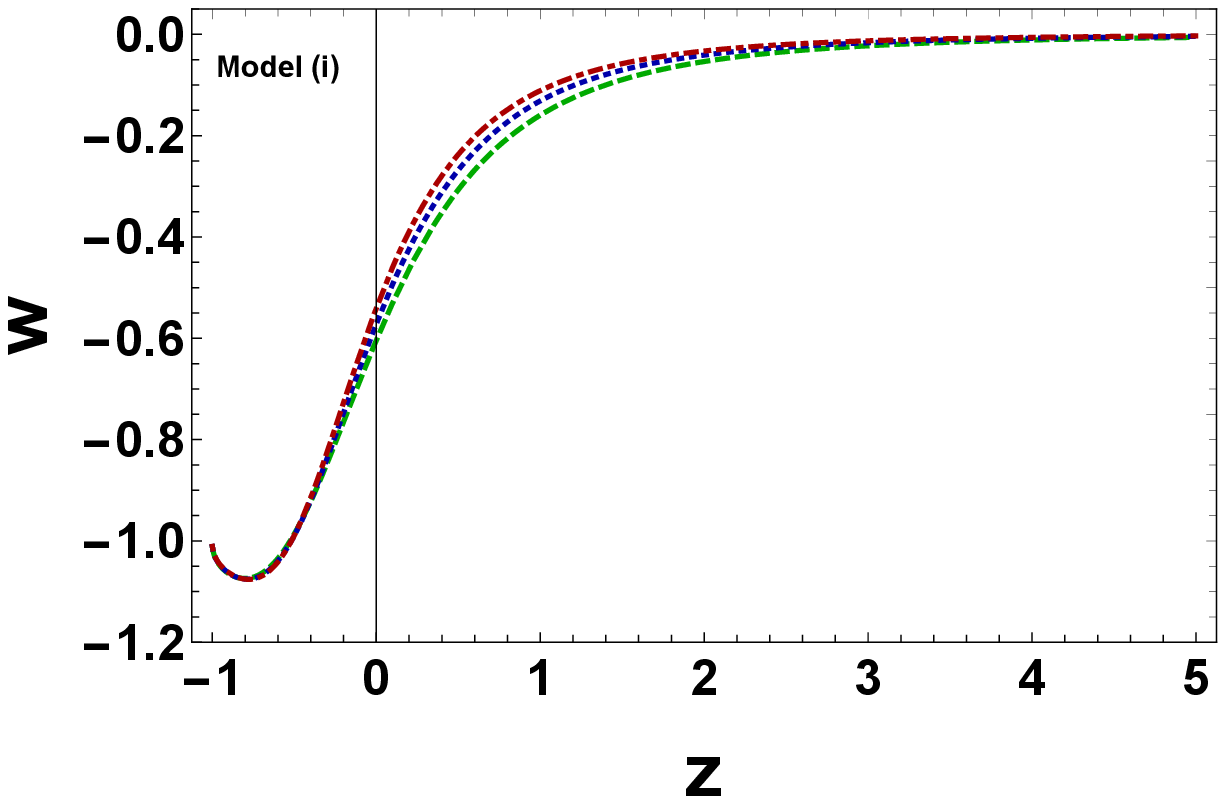}
\includegraphics[width=8cm]{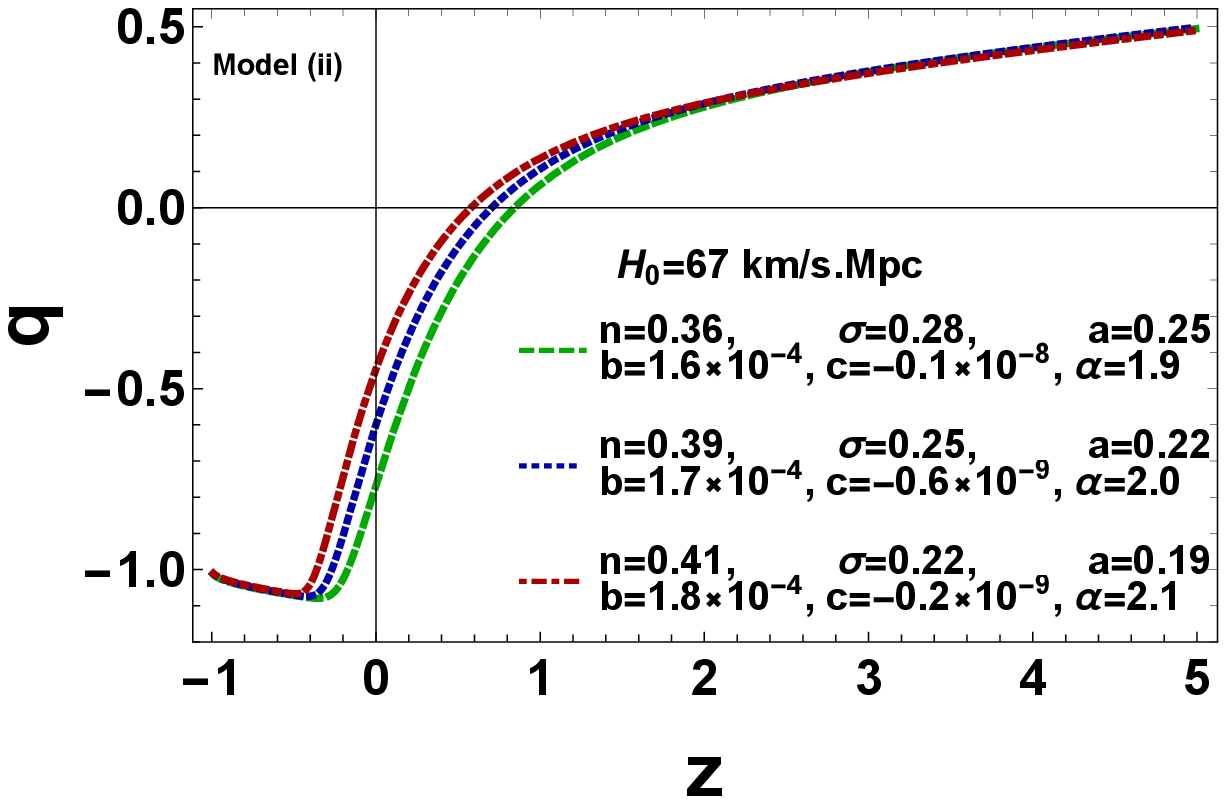}
\includegraphics[width=8cm]{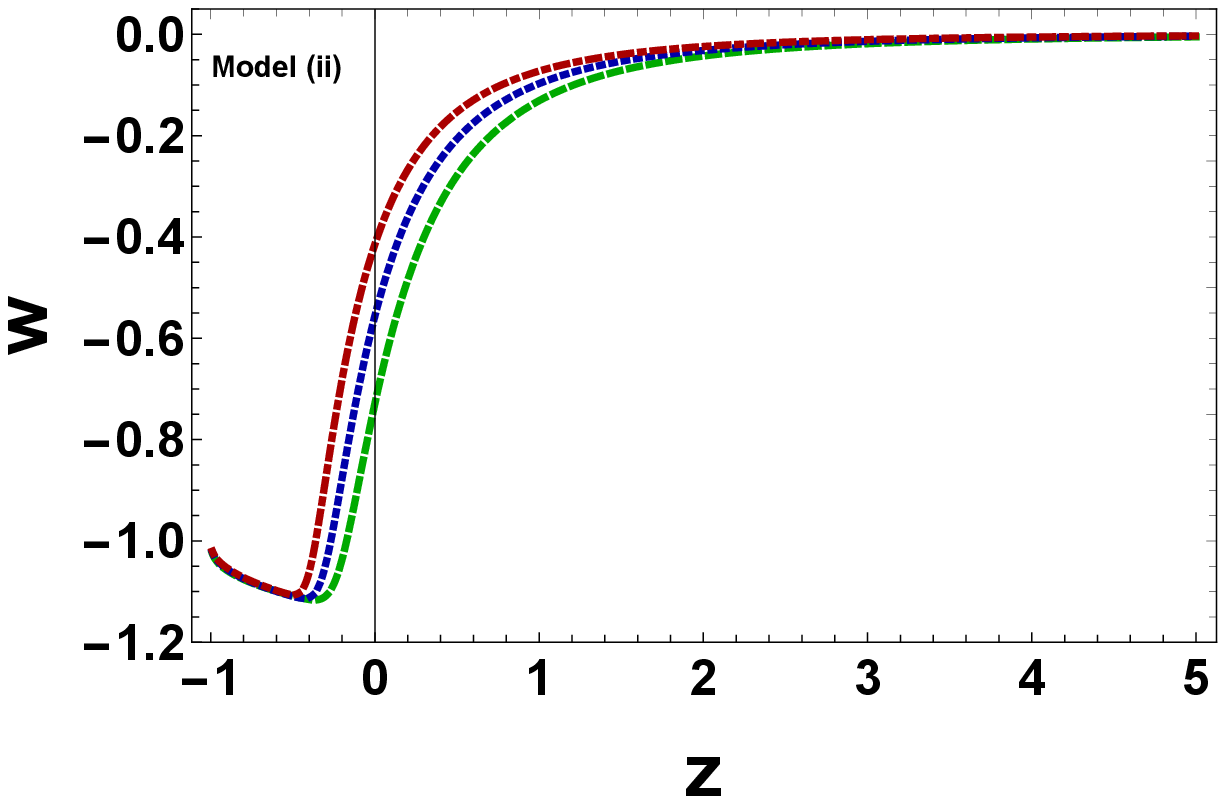}
\caption{The evolution of $q$ and state parameter $w$ versus $z$
for $H(z=0)=67$~\cite{dom}. Upper panels are provided for the case
(i) and the lower ones are depicted for case (ii) discussed in
Sect.~\ref{sec4}. {The model parameters used to draw the curves of $w$ are the same as those of $q$ diagrams.}}\label{1}
\end{center}
\end{figure}

The case with $f=a^{-n}$ leads to a decreasing function of time
whenever $n>0$~\cite{gaz,smo}. In this manner, assuming $w\equiv
p/\rho=0$, {together with} using Eqs.~(\ref{ras1}), one easily {finds}
$q=(3\alpha-1)^{-1}$, and $\rho=\rho_0 a^n(1-3\alpha
a)^{-(n+2)/n}$, where $\rho_0$ is the integration constant. These
results indicate that, at limit $a\rightarrow1$, the obtained
pressureless fluid can accelerate the universe expansion with
$q\leq-1/2$ for $-1/3\leq\alpha<1/3$. Consequently, the
non-minimal coupling $\gamma=[(1+z)^n-\alpha]/[4(1+z)^n-6\alpha]$
allows $G$ to vary as $G=G_0(1+z)^n$~\cite{gaz}, where we used the
$1+z=1/a$ relation. It is also easy to see that {the universe described by this model} has
begun from a primary inflationary phase ($q=-1$) corresponding to
the $a\rightarrow0$ point. In fact, in this limit, we also have
$\gamma=1/4$, a value that supports an inflationary phase for even
an empty universe~\cite{mor1}.

Now, let us consider two more comprehensive cases i.e., $i$)
$p=k\rho^{1+1/m}$, where $m$ and $k$ are unknown constants
{ to be} evaluated later, and $ii$) $p=\sigma\rho/(a-b\rho)-c\rho^2$ in
which $\sigma$, $a$, $b$ and $c$ are unknown coefficients. In this
manner, as it is obvious from Fig.~\ref{1}, { a} proper behavior is
obtainable for {the} cosmos. Here, $w\equiv p/\rho$ denotes the equation
of state of cosmic fluids. Depending on the values of unknown
parameters, the universe can also experience a transition at $z_t$
which can even take values smaller than $1$. Clearly, both fluids
behave as dark energy sources, and the corresponding non-minimal
coupling can not be considered as a dark energy source.
%%%%%%%%%%%%%%%%%%%%%%%%%%%%%%%%%%%%%%%%%%%%%%%%%%%
\section{A new proposal for $\lambda$ parameter}\label{sec5}

Now, let us consider a flat FRW universe filled by a pressureless
fluid with energy density $\rho$ when $\lambda R=\zeta H^n$ in
which $\zeta$ and $n$ are unknown constants. In this manner, 
the $\lambda$ parameter takes the form
\begin{eqnarray}
\lambda=\zeta\frac{H^n}{R}=\frac{\zeta}{6}\frac{H^n}{\dot{H}+2H^2},
\end{eqnarray}
whence, the corresponding Friedmann equations read
\begin{eqnarray}\label{fri}
&H^2-\frac{\kappa\zeta}{3}H^n=\frac{\kappa}{3}\rho,\nonumber\\
&H^2+\frac{2}{3}\dot{H}-\frac{\kappa\zeta}{3}H^n=0.
\end{eqnarray}

\begin{figure}[ht!]
\begin{center}
\includegraphics[width=8cm]{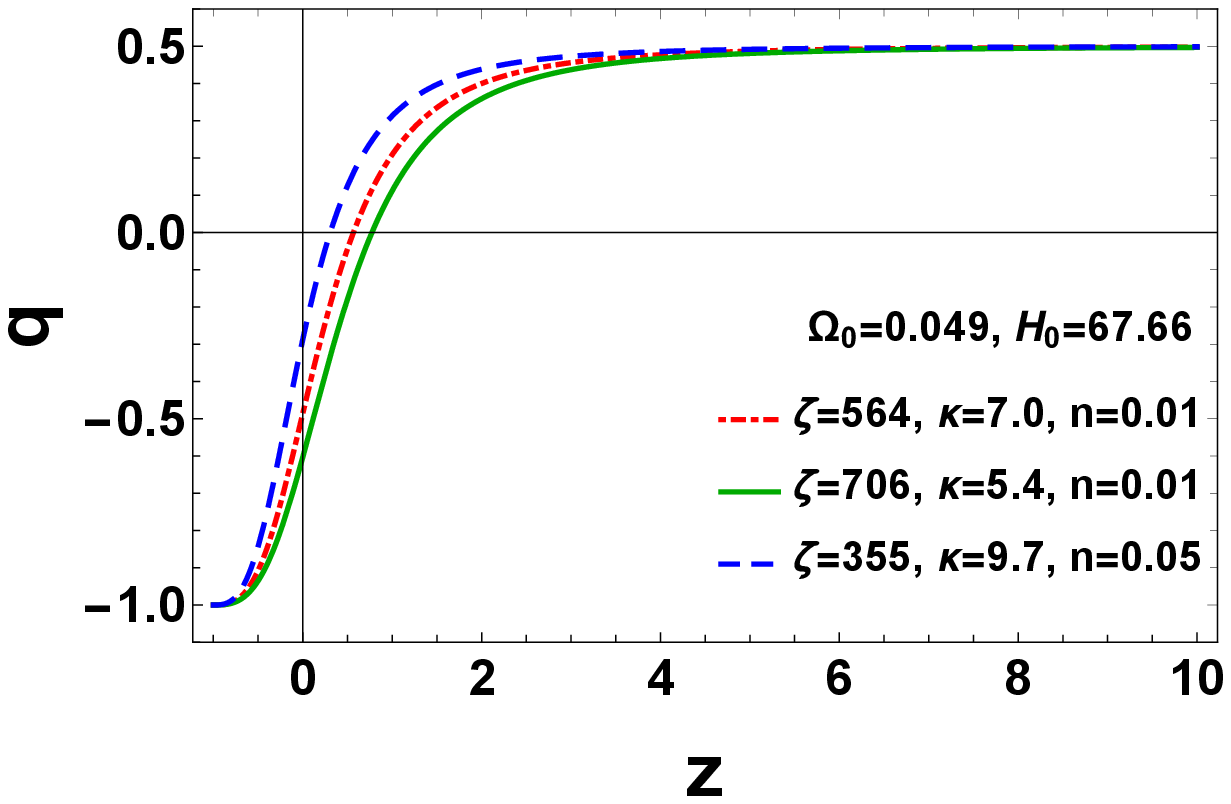}
\includegraphics[width=8cm]{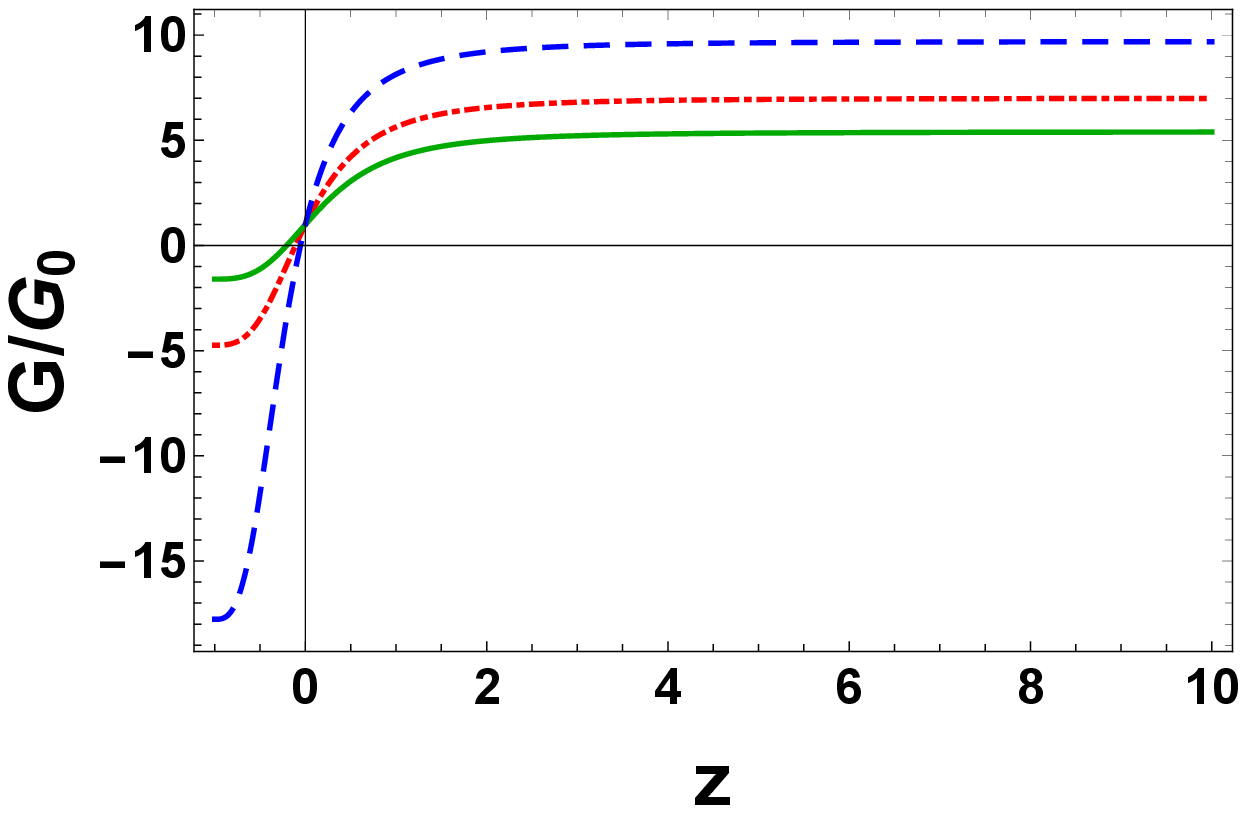}
\caption{The evolution of $q$ and $G/G_0$ assuming $w=0$,
for the case discussed in Sec.\ref{sec5}. {The diagrams for $G/G_0$ are plotted using the same model parameters as of $q$ diagrams}.}\label{3}
\end{center}
\end{figure}

\begin{figure}
\begin{center}
\includegraphics[width=8.2cm]{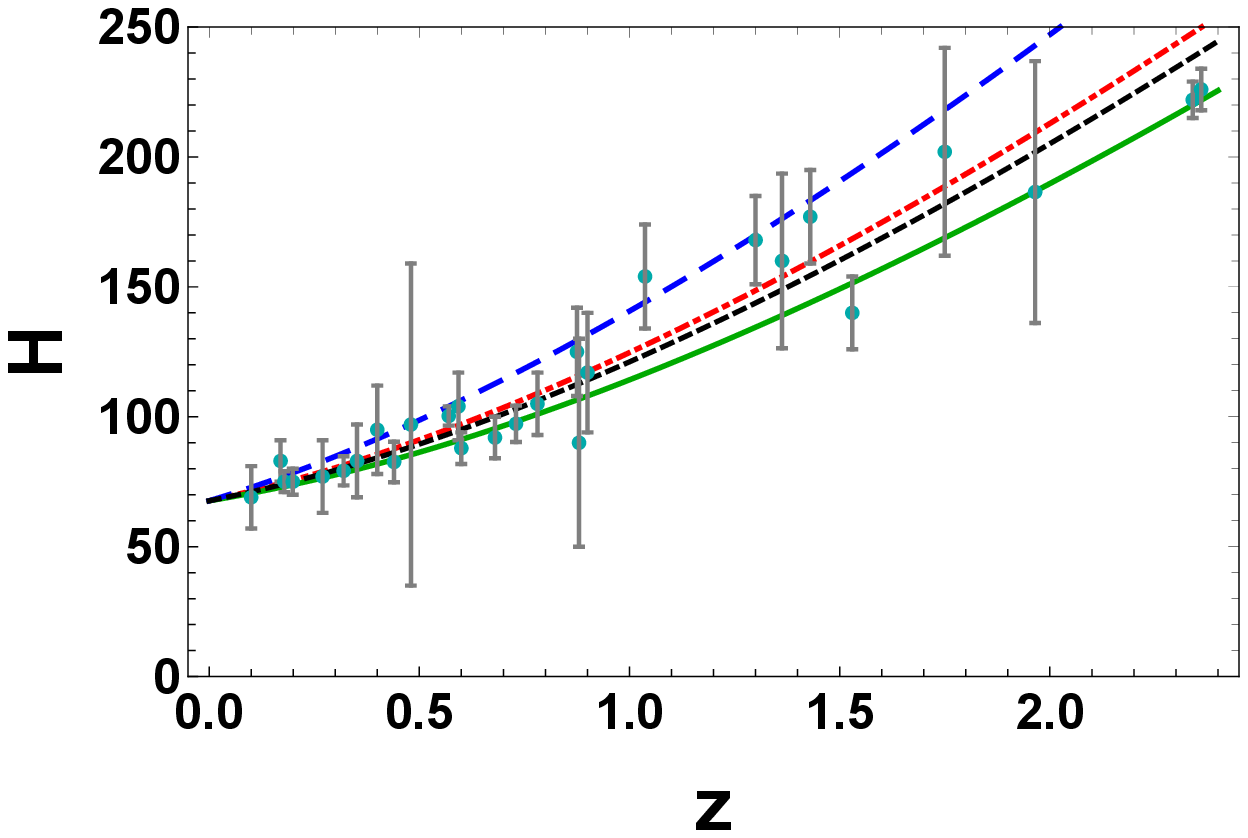}
\includegraphics[width=8cm]{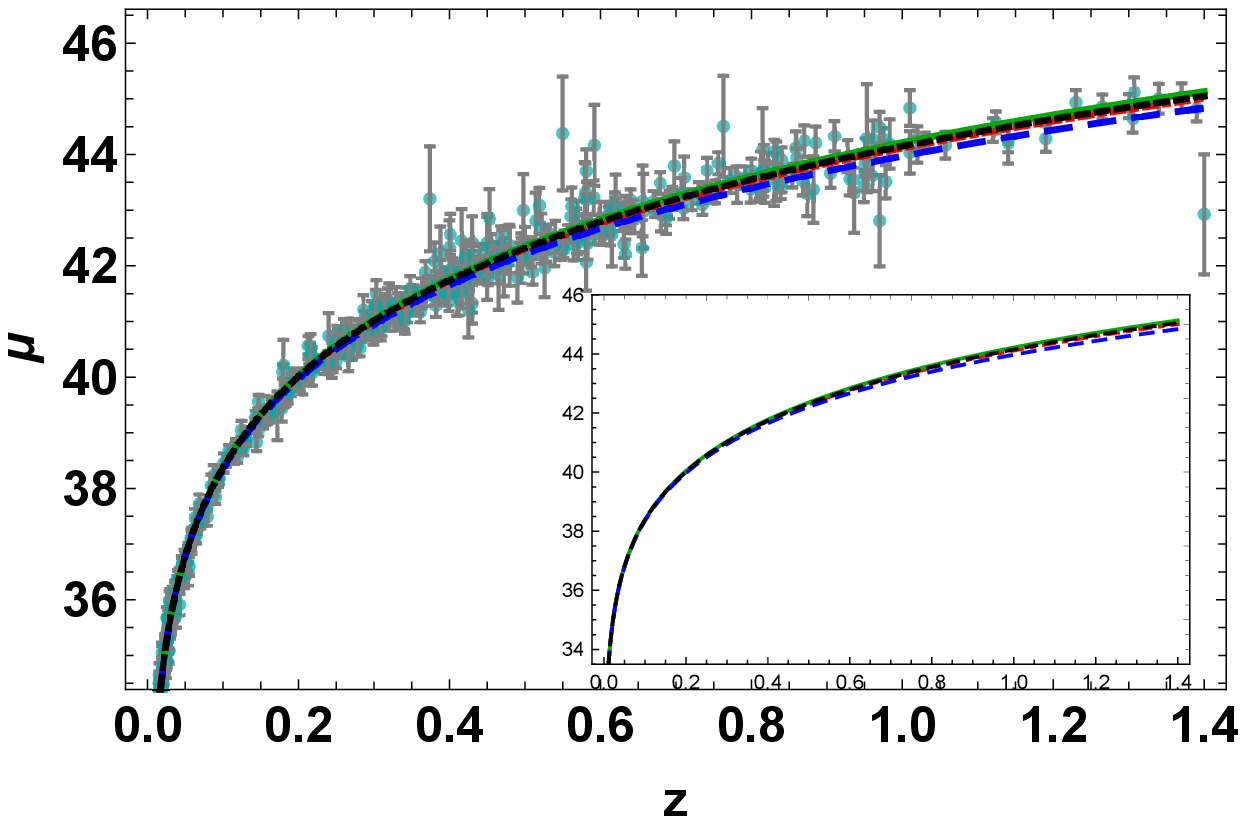}
\caption{The evolution of $H(z)$ and $\mu(z)$ whenever $w=0$, for
the case discussed in Sec.\ref{sec5}. {The same values of parameters as of Fig.~\ref{3} have been used. The black dashed lines show $H(z)$ and $\mu(z)$ for $\Lambda$CDM model.}}\label{4}
\end{center}
\end{figure}

\noindent Defining $\Omega=8\pi G\rho/3H^{2}$, while $\Omega_0$
denotes its current value~\cite{macq}, the evolution of $q$ and
$G/G_0$ have been plotted in Fig.~(\ref{3}). For the employed
parameters, transition redshift ($z_t$) lies within the range of
$0.4\leq z_t\leq0.88$. The sensitivity of diagrams to the values
of $\Omega_0$ and $H_0$ is so weak compared with those of $\zeta$
and $n$ and $\kappa$. Indeed, although we only consider a baryonic
source for current density parameter $\Omega_0=0.049$
~\cite{macq}, and $H_0=67.66$~\cite{agh}, the obtained behaviors
are also achievable for other candidates of $\Omega$ (such as dark
matter) and also the other values of $H_0$, reported in the
literature. Hence, suitable behavior of $q$ is obtainable by only
considering the baryonic content of the universe, meaning that the
$\zeta H^n$ term may play the role of the unknown parts (dark
components) of cosmos. Dirac hypothesis is also respected during
the cosmic evolution. Remarkably, $G$ will take negative values in
future meaning that gravity will become repulsive which speeds the
universe expansion rate up more i.e., $q$ decreases. All these
happen under the shadow of the existence of non-minimal coupling
$\lambda$ { which varies during the evolution of the universe}.
In Fig.~(\ref{4}), $H(z)$~\cite{far} and the distance
modulus~\cite{ama} are plotted for the $\Lambda$CDM model and also
our model.

The negative value of $G$ is the direct result of the assumed
$\lambda$, and changes in the values of model parameters do not
affect this result. There are also other works that predict
negative values for $G$ ~\cite{bar2,uza1,uza2}. Theoretically, our
model shows that a non-minimal coupling between geometry and
matter fields can accelerate the universe expansion and has an
ability to satisfy Dirac hypothesis.
%%%%%%%%%%%%%%%%%%%%%%%%%%%%%%%%%%%%%%%%%%%%%%%%%%%%%%%%
\section{concluding remarks}\label{sec6}
After addressing some properties of previously cosmological models
introduced in the framework of GRT ~\cite{mor1,das}, the
implications of GRT on obtaining varying $G$ has been studied
through considering the Newtonian limit of the field equations.
Thereinafter, following a proposal of Dirac hypothesis introduced
in~\cite{gaz,smo}, the required non-minimal coupling needed to
support Dirac model was also obtained. Our results show that the
dark sectors of cosmos can be unified into one cosmic fluid which
behaves as a pressureless fluid in high redshift limit, and also
accelerates the universe in line with the current
observations~(Fig.~\ref{1}). We also proposed a non-minimal
coupling~(Sec.~\ref{sec5}) which can play the role of dark side of
cosmos {satisfying} Dirac hypothesis. Indeed, the present
study addresses a deep connection between non-minimal coupling
(between the matter fields and geometry) and the idea of variable
$G$. { This translates into saying that} one may find the { footprints of} non-minimal coupling between the
matter {fields} and geometry by having the observationally
confirmed profile of $G$ and conversely.
\par
Although relying on the Rastall hypothesis { on} relation
between the changes in spacetime curvature and violation of
OCL we only focused on the implications of the violation of OCL
{ in} cosmology and its connection with Dirac hypothesis, the OCL
violation can also be allowed due to the quantum considerations
such as uncertainty principle, and in the framework of unimodular
gravity producing significant cosmological outcomes~\cite{jos}.
Indeed, even in the framework of GR and thanks to the Bianchi
identity, OCL is violated as the result of the existence of a
non-constant $G$. In summary, it was our goal to address $i$)
probable connection between Dirac hypothesis and non-minimal
couplings, and simultaneously, $ii$) the ability of such couplings
in being responsible for the unknown parts (dark sides) of cosmos.
Therefore, such couplings need to be further studied from both of
the theoretical and observational viewpoints. { Finally, we would like to mention that, though Rastall gravity and its generalizations provide interesting results, cosmological models based on this theory need to be accurately tested by observations. In the present model, we tried to explore theoretical consequences of a varying G cosmology based on GRT and also briefly examined observational aspects of the theory. However, a full observational treatment of the present model, e.g., in light of~\cite{Akarsu2020}, needs to be done and work along this line can be considered as an interesting subject for future studies and developments.}
%%%%%%%%%%%%%%%%%%%%%%%%%%%%%%%%%%%%%%%%%%%%%%%%%%%%%%%%
%\acknowledgments{The authors would like to appreciate the anonymous referee(s) for providing useful comments and suggestions that helped us to improve the original version of our manuscript.}
%%%%%%%%%%%%%%%%%%%%%%%%%%%%%%%%%%%%%%%%%%%%%%%%%%%%%%%


\begin{thebibliography}{99}
\bibitem{dir1} Dirac P. A. M., \textit{Nature} \textbf{139} (1937), 323
\bibitem{dir2} Dirac P. A. M., \textit{ Proc. Roy. Soc. London, Ser. A} \textbf{165} (1938), 199.
\bibitem{dir3} Dirac P. A. M., {\it Nature}, {\bf 139}, (1937), 1001.
\bibitem{WZE} Weyl H. {\rm Ann. Phys.}, {\bf 59}, 129 (1919);\\ Zwicky F., {\it Phys. Rev.}, {\bf 55}, 726 (1939);\\Eddington A. S., {\it The Mathematical Theory of Relativity,} Cambridge University Press, London (1923).
\bibitem{BTB} Barrow J. D., Tipler F. J., {\it The Anthropic Cosmological Principle}, Oxford University Press, Oxford, (1986);\\ J. D. Barrow, {\it Varying G and Other Constants,} In: S$\acute{a}$nchez N., Zichichi A. (eds) Current Topics in Astrofundamental Physics: Primordial Cosmology. NATO ASI Series (Series C: Mathematical and Physical Sciences), vol 511. Springer, Dordrecht (1998).
\bibitem{CHK} Chandrasekhar S., {\it Nature} {\bf 139} (1937), 757;\\Kothari D. S., {\it Nature}, {\bf 142} (1938), 354.
\bibitem{DIRACREV} S. Ray, U. Mukhopadhyay, S. Ray, A. Bhattacharjee,  {\it Int. Journal Mod. Phys. D} {\bf 28} (2019), 1930014.
\bibitem{ras2} Rastall P., \textit{Can. J. Phys.} \textbf{54} (1976), 66
\bibitem{vin} Vinti J. P., \textit{Celestial Mechanics} \textbf{16} (1977), 391
\bibitem{sab} De Sabbata V., \textit{Acta Cosmologica} \textbf{Zesz. 9} (1980), 63
\bibitem{bap} Baptista J. P., Batista A. B., Fabris J. C., \textit{Revista Brasileira de Fisica.} \textbf{14} (1984), 208
\bibitem{bee} Beesham A., \textit{Int. J. Theo. Phys.} \textbf{25} (1986), 1295
\bibitem{wu} Wu Y. S., Wang Z., \textit{Phys. Rev. Lett.} \textbf{57} (1986), 16
\bibitem{deg} Degl'Innocenti S. et al., \textit{A}\&\textit{A} \textbf{312} (1996), 345
\bibitem{bar1} Barrow J. D., \textit{Mon. Not. R. Astron. Soc.} \textbf{282} (1996), 1397
\bibitem{bar2} Barrow J. D., 1997, arXiv:gr-qc/9711084.
\bibitem{bar3} Barrow J. D., {\it The Constants of Nature}, (Vintage Books, London, 2002)
\bibitem{mans} Mansouri R., Nasseri F., Khorrami M., \textit{Phys. Lett. A} \textbf{259} (1999), 194
\bibitem{gaz} Gazta\~{n}aga E. et al., \textit{Phys. Rev. D} \textbf{65} (2001), 023506
\bibitem{clif} Clifton T., Mota D., Barrow J. D, \textit{Mon. Not. R. Astron. Soc.} \textbf{358} (2005), 601
\bibitem{bro} Bronnikov K. A., Kononogov S. A., \textit{Metrologia} \textbf{43} (2006), 1
\bibitem{sol} Sol\`{a} J.,  \textit{ J. Phys. A: Math. Theor.} \textbf{41} (2008), 164066
\bibitem{uza1} Uzan J. P., \textit{Rev. Mod. Phys.} \textbf{75} (2003), 403
\bibitem{uza2} Uzan J. P., \textit{Liv. Rev. Relativ.} \textbf{14} (2011), 2
\bibitem{smo} Smolin L., \textit{Class. Quantum Grav.} \textbf{33} (2016), 025011
\bibitem{fri} Fritzsch H., Sol\`{a} J., Nunes R. C., \textit{Eur. Phys. J. C} \textbf{77} (2017), 193
\bibitem{les} Leszczy\'{n}ska K., D\c{a}browski M. P., Denkiewicz T., \textit{Eur. Phys. J. C} \textbf{79} (2019), 222
\bibitem{ell} Ellis G. F. R., Maartens R., Maccallum M. A. H., {\it The Constants of Nature}, (Cambridge University Press, UK, 2012).
\bibitem{CanutoAdams} Canuto, V., Adams, P. J., Hsieh, S. H., Tsiang, E., Phy. Rev. D {\bf 16}, 6 (1977);\\Wesson, P., Goodson, R. E., Observ. {\bf 101}, 105 (1981).
\bibitem{11} C. Brans, R. H. Dicke, Phys. Rev. {\bf 124}, 925 (1961).
\bibitem{12} R. H. Dicke, Phys. Rev. {\bf 125}, 2163 (1962).
\bibitem{13} R. H. Dicke, Rev. Mod. Phys. {\bf 29}, 355 (1957).
\bibitem{14} R. H. Dicke, Nature {\bf 192}, 440 (1961).
\bibitem{pois} Poisson E., {\it A Relativist's Toolkit}, (Cambridge University Press, UK, 2004).
\bibitem{od1} Nojiri S., Odintsov S. D., \textit{Phys. Lett. B} \textbf{599} (2004), 137
\bibitem{all} Allemandi G. et al., \textit{Phys. Rev. D} \textbf{72} (2005), 063505
\bibitem{koi} Koivisto T.,\textit{Class. Quant. Grav.} \textbf{23} (2006), 4289
\bibitem{bert} Bertolami O. et al., \textit{Phys. Rev. D} \textbf{75} (2007), 104016.
\bibitem{hark} Harko T., Lobo F. S. N., \textit{Galaxies} \textbf{2} (2014), 410.
\bibitem{car} Carloni S., \textit{Phys. Lett. B} \textbf{766} (2017), 55.
\bibitem{boh} Boehmer C. G., Carloni S.,\textit{Phys. Rev. D} \textbf{98} (2018), 024054.
\bibitem{ras1} Rastall P., \textit{Phys. Rev. D} \textbf{6} (1972), 3357.
\bibitem{mor1} Moradpour H. et al., The European Physical Journal C, \textbf{77} (2017), 259.
\bibitem{mora} De Moraes W. A. G., Santos A. F., \textit{Gen. Relativ. Gravit.} \textbf{51} (2019), 167.
\bibitem{li} Li R. et al., \textit{Mon. Not. R. Astron. Soc.} \textbf{486} (2019), 2407.
\bibitem{raw1} Al-Rawaf A. S., Taha O. M., \textit{Phys. Lett. B} \textbf{366} (1996), 69.
\bibitem{raw2} Al-Rawaf A. S., Taha O. M., \textit{Gen. Relat. Gravit.} \textbf{28} (1996), 935.
\bibitem{raw3} Al-Rawaf A. S., \textit{Int. J. Mod. Phys. D} \textbf{14} (2005), 1941.
\bibitem{maj} Majernik V., \textit{Gen. Relat. Gravit.} \textbf{35} (2003), 1007.
\bibitem{arb} Arbab A. I., \textit{J. Cosmol. Astropart.  Phys.} \textbf{05} (2003), 008.
\bibitem{rah1} Abdel-Rahman A. M. M., \textit{Astrophys. Space. Sci.} \textbf{278} (2001), 383.
\bibitem{rah2} Abdel-Rahman A. M. M., Hashim M. H. A., \textit{Astrophys. Space. Sci.} \textbf{298} (2005), 519.
\bibitem{rah3} Abdel-Rahman A. M. M., Riad I. F., \textit{Astron. J.} \textbf{134} (2007), 1931.
\bibitem{mor2} Moradpour H. et al., \textit{Phys. Rev. D.} \textbf{96} (2017), 123504.
\bibitem{man} Manna T., Rahaman F., Mondal M., \textit{Mod. Phys. Lett. A} \textbf{35} (2020), 2050034.
\bibitem{ortiz2020} S. K. Maurya and F. T.-Ortiz, Phys. Dark Univ. {\bf 29} (2020), 100577.
\bibitem{shabooni2020} H. Shabani and A. H. Ziaie, Europhysics Letters {\bf 129}, (2020) 20004.
\bibitem{fab} Fabris J. C., Kerner R., Tossa J., \textit{ Int. J. Mod. Phys. D} \textbf{9} (2000), 111.
\bibitem{jos} Josset T., Perez A., \textit{ Phys. Rev. Lett. 118} \textbf{118} (2017), 021102.
\bibitem{wat} Watson S. et al., \textit{J. Cosmol. Astropart.  Phys.} \textbf{07} (2017), 11.
\bibitem{gam} Gamboa J. et al., \textit{Phys. Rev. D} \textbf{96} (2017), 083534.
\bibitem{das} Das D., Dutta S., Chakraborty S., \textit{Eur. Phys. J. C} \textbf{78} (2018), 810.
\bibitem{lin} Lin K., Qian W. L., \textit{Eur. Phys. J. C} \textbf{80} (2020), 561.
\bibitem{batis} C. E. M. Batista, M. H. Daouda, J. C. Fabris, O. F. Piattella, D. C. Rodrigues, Phys. Rev. D {\bf 85}, (2012), 084008.
\bibitem{mor3} Moradpour H. et al., \textit{Mod. Phys. Lett. A} \textbf{32} (2017), 1750078
\bibitem{mor4} Moradpour H. et al., \textit{Adv. High Energy Phys.} \textbf{2018} (2018), 7124730
\bibitem{mor5} Moradpour H., Sadeghnezhad N., Hendi S. H., \textit{Can. J. Phys.} \textbf{95} (2017), 1257
\bibitem{ref} T. R. P. Carames. et al., Eur. Phys. J. C74 (2014) 3145.
\bibitem{dar} Darabi F. et al., \textit{Eur. Phys. J. C} \textbf{78} (2018), 25
\bibitem{mor6} Moradpour H. et al., \textit{Mod. Phys. Lett. A} \textbf{33} (2019), 1950096
\bibitem{mot} Mota C.E., et al., arXiv:2007.01968.
\bibitem{AHH2020} A. H. Ziaie, H. Moradpour, H. Shabani, Eur. Phys. J. Plus {\bf 135} (2020), 916.
\bibitem{Lohiya1996} D. Lohiya, A. Batra, S. Mehra, S. Mahajan and A. Mukherjee, Astron. Astrophys. Trans. {\bf 14} (1997), 199.
\bibitem{Nunez2019} I. Ayuso, J. P. Mimoso and N. J. Nunes, Galaxies, {\bf 7} (2019), 38.
\bibitem{Batista2001} A. B. Batista, J. C. Fabris and S. V. B. Goncalves, {\it Class. Quant. Grav.} {\bf 18} (2001), 1389.
\bibitem{Star1981} A. A. Starobinskij, Pisma v Astronomicheskii Zhurnal, {\bf 7}, (1981), 67; Soviet Astronomy Letters, {\bf 7}, (1981), 36, Translation.
\bibitem{Lin2020} K. Lin and W.-L. Qian, Eur. Phys. J. C {\bf 80} (2020), 561.
\bibitem{dom} Dom\'{i}nguez A. et al., \textit{Astrophys. J.} \textbf{885} (2019), 137.
\bibitem{macq} Macquart J. et al., \textit{Nature} \textbf{581} (2020), 391.
\bibitem{far} Farooq O. et al., \textit{Astrophys. J.} \textbf{835} (2017), 26.
\bibitem{agh} Aghanim N. et al., \textit{A}\&\textit{A} \textbf{641} (2020), A6.
\bibitem{ama} Amanullah et al., \textit{Astrophys. J.} \textbf{716} (2010), 712.
\bibitem{Akarsu2020} O. Akarsu, N. Katirci, S. Kumar, R. C. Nunes, B. Ozturk and S. Sharma, Eur. Phys. J. C {\bf 80} (2020), 1050.
\end{thebibliography}
\end{document}